\documentclass[preprint,showpacs,preprintnumbers,amssymb]{revtex4}

\usepackage{graphicx}
\usepackage{bm}

\usepackage{color}

\begin{document}

\title{A tentative geometrical description of static dilatancy in liquid foams: ordered 2D and 3D foams}

\author{Pierre Rognon~$^{1,2,4}$}
\email{prognon@usyd.edu.au}
\author{Fran\c{c}ois Molino~$^{3}$}
\email{francois.molino@igf.cnrs.fr}
\author{Cyprien Gay~$^{1}$}
\email{cyprien.gay@univ-paris-diderot.fr}

\affiliation{%
$^{1}$ Mati\`{e}re et Syst\`{e}mes Complexes, 
Universit\'{e} Paris Diderot--Paris 7, CNRS, UMR~7057, 
B\^atiment Condorcet, Case courrier 7056, F--75205 PARIS cedex 13, France\\
$^{2}$ Centre de Recherche Paul Pascal, CNRS, UPR~8641,
Universit\'e de Bordeaux~1,
115 Av. Schweitzer, F--33600 PESSAC, France\\
$^{3}$ Department of Endocrinology,
Institute of Functional Genomics, CNRS, UMR~5203,
INSERM U661, Universities of Montpellier~1 and~2,
141 rue de la Cardonille, F--34094 MONTPELLIER cedex 05, France\\
$^{4}$ Current address: 
School of Civil Engineering, J05, The University of Sydney, Sydney, New South Wales 2006, Australia.
}%

\date{September 25, 2009}


\begin{abstract}
Liquid foams have been observed to behave like immersed granular materials 
in at least one respect: deformation tends to raise their liquid contents, 
a phenomenon called dilatancy. 
We present a geometrical interpretation thereof
in foams squeezed between two solid plates (2D GG foams),
which contain pseudo Plateau borders along the plates,
and in 3D foams.
While experimental observations evidenced the effect 
of a continuous deformation rate (dynamic dilatancy), 
the present argument applies primarily 
to elastic deformation (static dilatancy).
We show that the negative dilatancy predicted 
by Weaire and Hutzler (Phil. Mag. 83 (2003) 2747)
at very low liquid fractions is specific to ideal 2D foams
and should not be observed in the dry limit of real 2D foams.
\end{abstract}

\maketitle

\newcommand{\hs}{\hspace{0.8cm}}
\newcommand{\be}{\begin{equation}}
\newcommand{\ee}{\end{equation}}
\newcommand{\bee}{\begin{eqnarray}}
\newcommand{\eee}{\end{eqnarray}}

\newcommand{\mysection}{\section}

\newcommand{\trace}{{\rm tr}}

\newcommand{\si}{\sigma}
\newcommand{\unittensor}{I}
\newcommand{\surfacetension}{\gamma}
\newcommand{\specificsurface}{\Sigma}
\newcommand{\siinterf}{\si^{\rm interf}}
\newcommand{\philiq}{\phi}
\newcommand{\deform}{\epsilon}

\newcommand{\dimratio}{k}
\newcommand{\truebidi}{{i-2d}}
\newcommand{\unstablebidi}{{u-2d}}

\newcommand{\pgas}{p_{\rm g}}
\newcommand{\pliq}{p_{\rm l}}
\newcommand{\Dp}{\Delta p}
\newcommand{\pwall}{p_{\rm wall}}
\newcommand{\pdis}{\Pi_{\rm d}}
\newcommand{\piosm}{\pi_{\rm osm}}
\newcommand{\dilatancy}{\chi}

\newcommand{\Sbub}{S} 
\newcommand{\Abub}{{\cal A}} 
\newcommand{\Vbub}{\Omega} 
\newcommand{\Vliq}{\Omega_{\rm liq}} 
\newcommand{\Vpb}{V_{\rm Pb}} 
\newcommand{\sppb}{S_{\rm hPb}} 
\newcommand{\Pbradius}{R} 
\newcommand{\pPbradius}{\Pbradius_{\rm ps}}  

\newcommand{\Vwn}{V_{\rm wn}} 
\newcommand{\shpb}{\sppb} 
\newcommand{\svpb}{S_{\rm vPb}} 

\mysection{Dilatancy in foams and grains}\label{Sec:Intro}

Liquid foams~\cite{weaire_hutzler_1999_book} 
and granular materials
are common examples of materials whose mechanical behaviours
exhibit complex features, such as non-homogeneous flows:
shear-banding, fracture or jamming.
In the present work, we are interested in yet another
such complex behaviour, called ``dilatancy''.
Dilatancy was described by Bagnolds~\cite{bagnold_1941}
in the context of granular materials:
upon deformation, because grains are forced
to move while avoiding each other,
the medium swells to some extent. 
In other words, the fluid volume fraction $\philiq$ is increased.
This effect can remain unnoticed in air.
By contrast, a spectacular absorption of liquid~\cite{bagnold_1941}
is obtained upon deformation of an immersed granular sample.

In liquid foams, it is not {\em a priori} so obvious 
whether dilatancy should be expected or not.
Because bubbles are deformable individually,
a foam should be able to deform substantially
without altering its liquid fraction.

In 3D foams, {\em dynamic} dilatancy 
was indeed observed under stationary shear~\cite{marze_2005_121}:
in a device where the shear rate localizes,
the liquid volume fraction was observed,
both visually and through electrical conductimetry,
to stabilize, within seconds, at a higher value
in the 
region being sheared continuously,
while it remained lower in both statically deformed regions
(and when shearing was stopped, the liquid fraction
was observed to become uniform again within seconds).
Dynamic dilatancy, whose microscopic origin
is yet uncertain~\footnote{It might originate 
in the slippage of Plateau borders along inter-bubble films,
and thus be related to the slippage 
of pseudo Plateau Borders along solid walls.
Experiments have been carried out in this 
direction recently~\cite{emile_2007_72,terriac_2006_909}.},
thus constitutes one out of many mechanisms
that could favour shear banding, 
a family of rheological behaviours
commonly observed in many other complex fluids
such as giant micelle solutions,
multilamellar vesicles or onions, granular pastes, etc.
The stabilizing mechanism is as follows.
As both the foam shear modulus
and its yield strain are known to decrease
as the liquid fraction increases~\cite{weaire_hutzler_1999_book},
a region that contains more liquid
will host statistically more plastic events than other regions
and, if dynamic dilatancy is indeed present,
even more liquid will then permeate towards that region
from less active regions.
This process reaches a stationary regime (co-existing shear bands)
when the dilatancy-induced pressure gradient
is balanced by the osmotic gradient 
associated with fluid concentration gradients.

Whether {\em static} dilatancy 
(resulting from a {\em fixed deformation})
also exists in liquid foams
has not been determined experimentally so far.
It might constitute a destabilizing factor
for a homogeneous flow and {\em generate} shear banding in the first place.

\begin{figure}
\begin{center}
\resizebox{10.0cm}{!}{%
\includegraphics{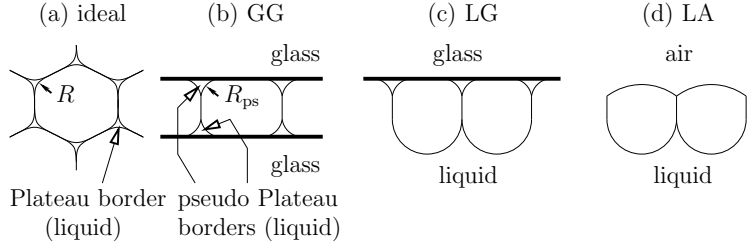}
}
\end{center}
\caption{Four types of 2D foams.
(a) Top-view of an ideal, truly two-dimensional foam:
either the third dimension is purely absent,
or the foam is invariant with respect to the third dimension.
(b) Side-view of a monolayer of bubbles 
between two {\em glass} plates (``GG'' foam).
(c) Monolayer of bubbles floating on a {\em liquid} bath
below a {\em glass} plate (``LG'' foam).
(d) Monolayer of bubbles floating 
on the free surface of the {\em liquid} bath,
in contact with {\em air} (``LA'' foam).
Dilatancy is no real issue for LG or LA foams
because of the underlying reservoir of liquid.
By contrast, the GG foam holds liquid not only in the Plateau
borders just like an ideal 2D foam
(second term of Eq.~(\ref{Eq:Vliq_PR})),
but also in the pseudo Plateau borders
which run along the edge between two bubbles
and touch one solid plate each
(first term of Eq.~(\ref{Eq:Vliq_PR})).
}
\label{Fig:2D_foam_types_GG_two_contributions}
\end{figure}

A {\em theory} of static dilatancy
was derived from thermodynamical considerations 
a few years ago~\cite{weaire_2003_2747,rioual_2005_117}
for an ideal, truly two-dimensional foam (Fig.~\ref{Fig:2D_foam_types_GG_two_contributions}a),
with predictions based on empirical or numerical results
concerning the osmotic pressure and the elastic modulus.
The model leads to the surprising prediction
that the sign of the dilatancy effect
should be negative for very low liquid fractions,
as a consequence of the fact that the elastic modulus
does not depend on the liquid fraction in the dry limit
(Decoration Theorem~\cite{bolton_weaire_1991_795,weaire_hutzler_1999_book}).

In this paper, we provide a simple geometrical description of static dilatancy in a dry liquid foam. 
An elastic deformation of such a material induces an elongation of the bubbles 
while conserving their volume constant,
thereby increasing the total Plateau border lengh per bubble, or per unit volume. 
Since most of the liquid is located in Plateau borders, the liquid fraction is thus increased.

According to this description, we calculate the magnitude of the dilatancy effect analytically 
for a crystalline, hexagonal GG foam 
(Fig.~\ref{Fig:2D_foam_types_GG_two_contributions}b) as well as for a 3D Kelvin foam, 
and we discuss negative dilatancy~\cite{weaire_2003_2747,rioual_2005_117}.



\mysection{Static dilatancy in a dry 2D glass-glass foam}
\label{Sec:dilatancy_2d_gg_foam}

Here, we introduce 
a very simple geometrical description of static dilatancy.
In a ``GG'' foam, there are two contributions
to the liquid content,
which are presented on Fig.~\ref{Fig:2D_foam_types_GG_two_contributions}:
\be
\label{Eq:Vliq_PR}
\Vliq \simeq
\frac{4-\pi}{2}\,P\,\pPbradius^2
+(2\sqrt{3}-\pi)\,\Pbradius^2 H
\ee
where $\Vliq$ is the volume of liquid per bubble.
The first term corresponds 
to the pseudo Plateau borders 
(with the corresponding radius of curvature $\pPbradius$
and the bubble perimeter $P$)
while the second term corresponds 
to the genuine Plateau borders 
(radius $\Pbradius$ and gap $H$ between the solid plates).

\begin{figure}
\begin{center}
\resizebox{10.0cm}{!}{%
\includegraphics{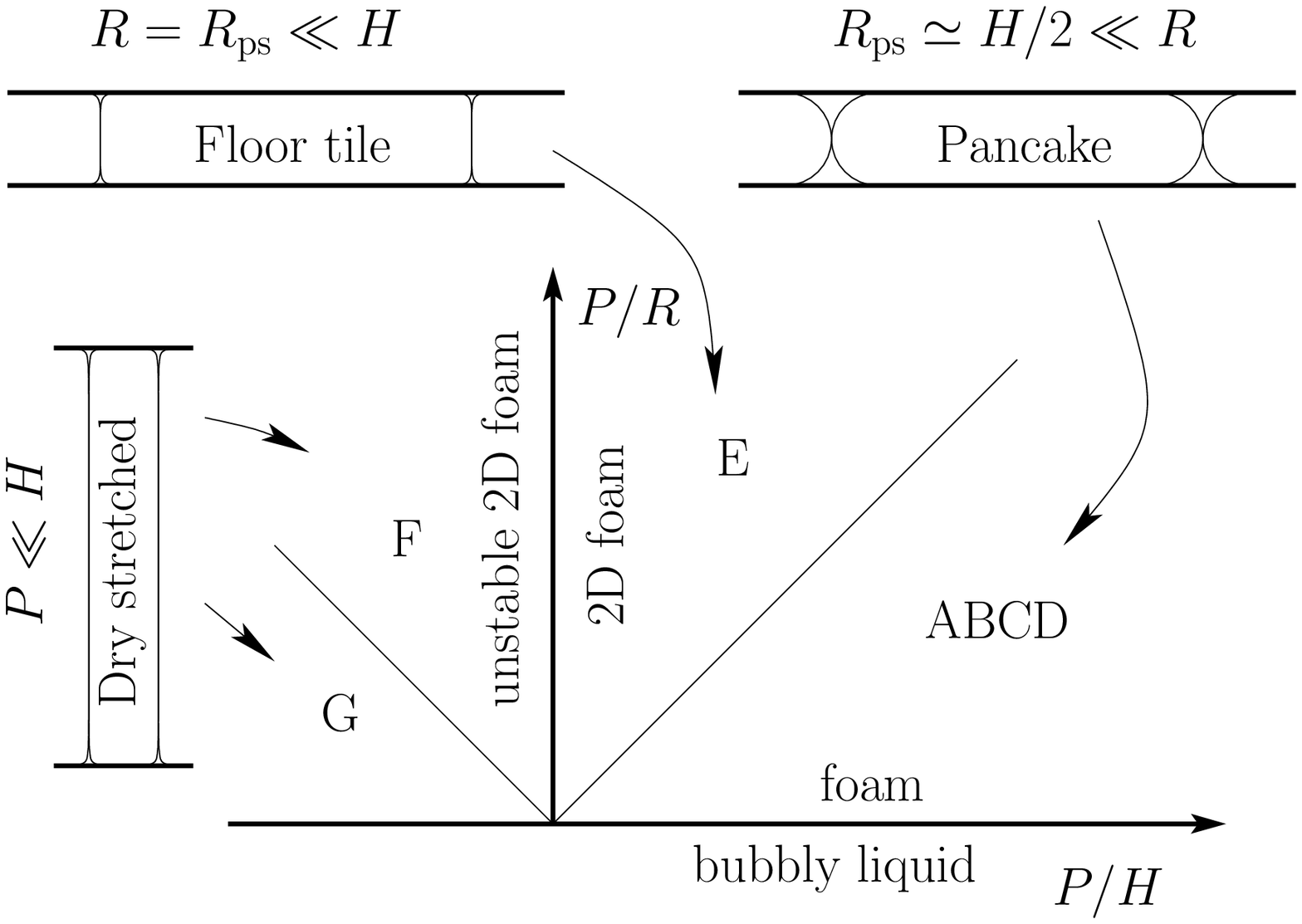}
}
\end{center}
\caption{Regimes of a GG-foam 
with low liquid fraction ($\philiq\ll 1$),
in terms of the bubble perimeter $P$ (measured on top view),
the Plateau border radius $\Pbradius$ (as seen from above too)
and the cell height $H$.
Real GG-foams ($P>H$) can be found in two main configurations:
the ``pancake'' regime~\cite{dilatancy_geometry_article}
($H-2\pPbradius\ll H\ll\Pbradius$, region ABCD)
and the ``floor tile'' regime E ($\Pbradius=\pPbradius\ll H$).
Regimes F and G do not correspond to real foams
(which cannot sustain the condition $H\gg P$
without destabilizing into 3D foams),
but regime G is the limit of truly two-dimensional foams ($H\rightarrow\infty$)
addressed in earlier works 
on dilatancy~\cite{weaire_2003_2747,rioual_2005_117}.}
\label{Fig:diagram_pancake_regimes_ABCD}
\end{figure}

In the present work, we will assume
both $P\gg R$ ({\em i.e.}, $\philiq\ll 1$)
and $\pPbradius=\Pbradius\ll H$ (floor tile regime).
In other words, we concentrate on regimes E, F and G
of Fig.~\ref{Fig:diagram_pancake_regimes_ABCD}.
Regime E corresponds to a floor tile GG-foam,
while regime G is the truly 2D limit 
of refs.~\cite{weaire_2003_2747,rioual_2005_117}

We here focus on floor tile ``GG'' foams
(regime E of Fig.~\ref{Fig:diagram_pancake_regimes_ABCD}),
{\em i.e.}, with not only a low liquid volume fraction
($\philiq\ll 1$, achieved when 
$\Pbradius\ll P$)
but also small pseudo Plateau borders
compared to the gap between the solid plates ($\pPbradius\ll H$),
in which case one has $\pPbradius\simeq\Pbradius$.
The opposite limit of ``pancakes'' 
($\pPbradius\simeq H/2\ll\Pbradius$, regime ABCD
of Fig.~\ref{Fig:diagram_pancake_regimes_ABCD})
will be discussed elsewhere~\cite{dilatancy_geometry_article}.

\begin{figure}
\begin{center}
\resizebox{10.0cm}{!}{%
\includegraphics{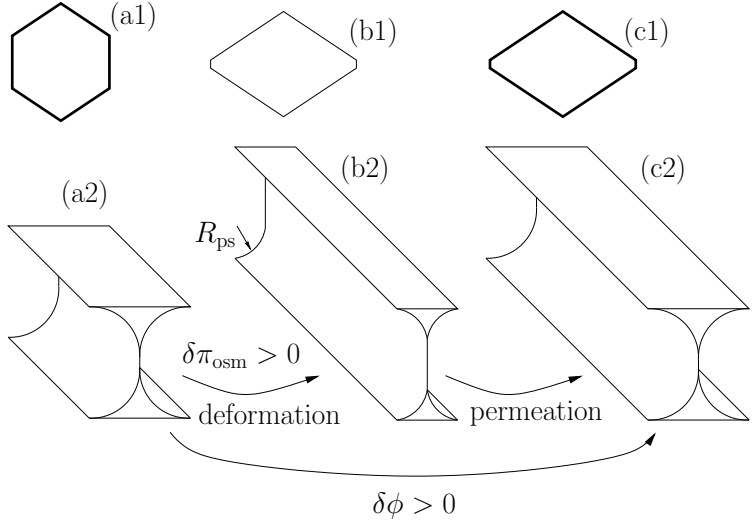}
}
\end{center}
\caption{Pseudo Plateau borders
between two bubbles squeezed between two glass plates.
Original configuration, with top-view (a1) and perspective (a2).
When a bubble which is initially at rest
is suddenly deformed, its perimeter, {\em i.e.,}
the total length of its pseudo Plateau borders, is increased.
On short time scales, 
the amount of liquid that surrounds the bubble remains constant,
hence the quantity of liquid per unit length
of the perimeter decreases (b1).
Thus, the pseudo Plateau borders shrink,
and the corresponding radius of curvature $\pPbradius$
is reduced (b2).
This implies that the pressure in the liquid is lowered.
As a result, liquid from less deformed regions of the foam,
where the liquid pressure is not as low,
can soon permeate and compensate for the Plateau border shrinking (c).
}
\label{Fig:elongation}
\end{figure}

Let us consider such a 2D glass-glass foam
free of any significant anisotropic ({\em i.e.}, deviatoric) in-plane stress.
If the foam is now deformed elastically,
the average bubble perimeter $P$ increases
as the typical bubble elongates
while conserving its volume constant.
Since the total amount of liquid per bubble 
remains constant on short time scales,
the increase in perimeter causes
the pseudo Plateau borders to shrink accordingly
(see Fig.~\ref{Fig:elongation}).
Later on, as permeation takes place,
the pseudo Plateau border radius $\pPbradius$
may return to its original value,
and the part of the liquid that is located 
in the pseudo Plateau borders increases accordingly.

Thus, soon after deformation,
the pressure difference between the gas and the liquid must increase
as the pseudo Plateau borders (and Plateau borders) shrink.
Later, after permeation has taken place,
the liquid content of the foam has increased
as compared to its value when the foam was at rest.
In other words, there is a positive dilatancy effect
as soon as the perimeter increases.

\begin{figure}
\begin{center}
\resizebox{10.0cm}{!}{%
\includegraphics{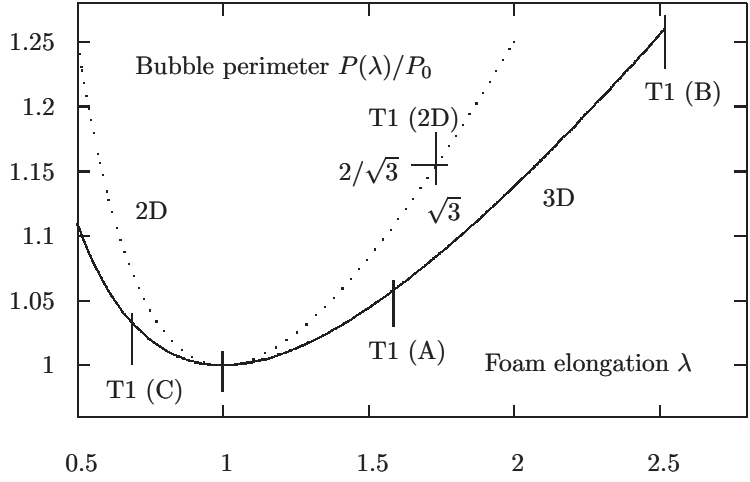}
}
\end{center}
\caption{Sum of all Plateau border lengths 
in a 2D (hexagonal, Eq.~(\ref{Eq:perimeter_dry})) foam
or 3D (Kelvin, Eq.~(\ref{Eq:perimeter_3d_AB})) foam,
as a function of the elongation $\lambda$.
In 3D case, elongation is performed in direction A 
(normal to a square facette),
B (normal to a hexagonal facette)
or C (one of the edges of Kelvin's cell)
with compression $1/\sqrt{\lambda}$
in both perpendicular directions.
}
\label{Fig:perimetre_2d_3d}
\end{figure}

In order to evaluate the change in bubble perimeter,
let us consider a dry, crystalline foam
subjected to an arbitrary elastic, homogeneous deformation.
Up to a global rotation, this deformation
is equivalent to an elongation by some factor $\lambda$
in one direction and to a compression by the same factor
in the perpendicular direction.
In such a hexagonal foam, in the dry limit,
the bubble perimeter increases in the following way:
\be
\label{Eq:perimeter_dry}
\frac{P^{\rm 2D}(\lambda)}{P^{\rm 2D}_0}=
\left[\frac{\lambda}{2}+\frac{1}{2\lambda}\right]
\hs
\frac{\delta P}{P^{\rm 2D}_0}
=\frac{(\lambda-1)^2}{2\lambda}
\ee
The perimeter thus increases
when the foam is deformed ($\lambda\neq 1$),
as can be seen on Fig.~\ref{Fig:perimetre_2d_3d}.
This law, which is exact in the dry limit, 
holds as long as no T1 process occurs
and does not depend on the orientation 
of the crystalline network
with respect to the direction of elongation.
In fact, it even holds for polydisperse
hexagonal foams~\cite{reinelt_kraynik_princen_1991_1235}.
The elongation at which T1 processes occur
depends both on orientation
or on liquid volume fraction.
In the dry limit, it is $\lambda=\sqrt{3}$
when elongation is perpendicular to some facets
and larger for other elongations.
This yield deformation is lowered as the foam becomes wetter
as described already long ago~\cite{weaire_hutzler_1999_book}.
The expected perimeter increase for $\lambda=\sqrt{3}$
is of order 15\%,
which means that gas-liquid pressure difference 
is expected to increase by 7\% on short time scales,
while the foam liquid fraction
is expected to increase by 15\% after permeation.

\mysection{Dilatancy in three dimensions}

In a dry, three-dimensional foam,
most of the liquid is located in the Plateau borders.
Following the same line of thought as above,
static dilatancy is expected
if deformation induces an increase
in the total Plateau border length per bubble
(or per unit volume of foam).

To test this in a simple manner,
we took an approximate version 
of Kelvin's bcc bubble packing.
Kelvin's packing is one of the most 
commonly encountered in monodisperse 3D foams,
even though it does not have the 
lowest energy~\cite{weaire_phelan_1994_107}.
In a real Kelvin foam,
square faces have edges bending slightly outwards,
while hexagonal faces are slightly non-planar.
Here, each bubble is assumed to be a truncated octahedron,
called {\em tetrakaidecahedron},
with six planar, square faces
and eight regular hexagonal, planar faces.
This approximation was used and discussed 
by Reinelt and Kraynik~\cite{reinelt_kraynik_1993_460}
to express the stress and the energy
under large elastic deformations.
Here, we present the changes in the total
Plateau border perimeter.

Computing the bubble perimeter
(sum of all edge lengths)
as a function of elongation
has been performed for three orientations of the elongation
(with a compression by $1/\sqrt{\lambda}$
in the perpendicular directions.
As can be seen on Fig.~\ref{Fig:perimetre_2d_3d},
the result is very similar to the two-dimensional case,
and indicates that significant dilatancy
is to be expected in dry 3D foams, too.
Incidentally, Fig.~\ref{Fig:perimetre_2d_3d}
indicates that for the present approximation
of a Kelvin foam, the response is 
independent of the direction of elongation:
\be
\label{Eq:perimeter_3d_AB}
\frac{P_A^{\rm 3D}(\lambda)}{P_0^{\rm 3D}}
=\frac{P_B^{\rm 3D}(\lambda)}{P_0^{\rm 3D}}
=\frac{P_C^{\rm 3D}(\lambda)}{P_0^{\rm 3D}}
=\frac13\left[\lambda+\frac{2}{\sqrt{\lambda}}\right]
\ee

\mysection{Osmotic pressure and dilatancy}

The {\em dilatancy coefficient} is defined~\cite{weaire_2003_2747}
as the (second) derivative of the osmotic pressure
with respect to the deformation $\epsilon$:
$\dilatancy=\partial^2\piosm/\partial\epsilon^2$.
%
With the variations of the average bubble perimeter in mind,
let us now derive the variation of the osmotic pressure 
in a floor tile GG-foam.
From Eq.~(\ref{Eq:perimeter_dry}), 
expressed in terms of $\lambda=1+\epsilon$,
one has $\delta P/P\simeq\epsilon^2/2$.
Hence, $\dilatancy=\delta\piosm/(\delta P/P)$,
where the variation $\delta\piosm$
is taken at constant liquid volume fraction $\philiq$.

When the foam is confined in a container,
the osmotic pressure $\piosm$
corresponds to the force that must be applied externally
to one of the confining walls if the latter
is permeable to the liquid but not to the bubbles.
The osmotic pressure (which is in fact a symmetric tensor
and not just a scalar quantity~\cite{weaire_2003_2747,rioual_2005_117})
is thus the difference between the stress in the foam
and the pressure applied by the pure liquid 
on the other side of the semi-permeable wall:
$\piosm=-\pliq-\si$
(where tensile stresses and pressures are both counted positively).
The stress in the foam 
includes a pressure contribution 
from the liquid ($\pliq$)
and from the gas ($\pgas$),
as well as a tensile contribution from the interfaces:
$\si=-\philiq\pliq\unittensor
-(1-\philiq)\pgas\unittensor
+\siinterf$. 
In the floor tile limit ($\Pbradius\ll H$), 
the interface stress,
averaged over in-plane orientations
and over the sample thickness, 
is related to the bubble perimeter~\cite{dilatancy_geometry_article}:
$<\siinterf>_{2D}\simeq(1-\philiq)\,\left[\frac{\surfacetension P H}{2\Vbub}
\right]+\frac{2\surfacetension}{H}$,
where $\surfacetension$ is the surface tension
and $\Vbub$ is the bubble volume.
Hence, osmotic pressure (now expressed as a scalar quantity) 
can be written as:
$<\piosm>_{2D}=(1-\philiq)\,
\left[\Dp-\frac{\surfacetension\,P\,H}{2\Vbub}
\right]-\frac{2\surfacetension}{H}$,
where
$\Dp=\pgas-\pliq= \frac{\gamma}{\pPbradius} = \frac{\gamma}{\Pbradius}$
is the pressure difference 
between the gas and the liquid.
The dilatancy then results 
from Eqs.~(\ref{Eq:Vliq_PR}) and~(\ref{Eq:perimeter_dry}):
\be
\label{Fig:dilatancy}
\dilatancy=\left.\frac{\delta<\piosm>_{2D}}{\delta P/P}\right|_\philiq
\simeq\frac{\surfacetension/(2\Pbradius)}
{1+\frac{\sqrt{3}-\pi/2}{1-\pi/4}\frac{H}{P}}
-\frac{\surfacetension\,P}{2\Abub}
\ee
The last term, in which $\Abub=\Vbub/H$, 
originates in the interface stress contribution $\siinterf$
and is equal to one half of the interfacial energy per unit height
($\surfacetension\,P$ per bubble).

\mysection{Negative dilatancy?}

For a GG-foam (regime E with $P\gg H$,
see Fig.~\ref{Fig:diagram_pancake_regimes_ABCD}),
one has $\dilatancy\simeq\Dp/2>0$.
By contrast, when $H\gg P$ (regime G), 
one gets the limit of a truly two-dimensional foam
since the pseudo Plateau borders (total length $P$ per bubble)
are too small compared to the Plateau borders (length $H$)
to contribute to dilatancy significantly.
Hence only the last term remains in Eq~(\ref{Fig:dilatancy}),
and dilatancy is negative:
$\dilatancy\simeq-\frac{\surfacetension\,P}{2\Abub}$.
We thus recover exactly the result $\dilatancy\simeq-2G$
by Weaire and Hutzler~\cite{weaire_2003_2747} in the
dry limit~\footnote{The shear modulus $G$ can be computed as follows.
The change in energy for a shear strain $\Gamma$
is $\delta{\cal E}=\frac12 G \Gamma^2$
and is also related to the change in the bubble perimeter:
$\delta{\cal E}=\surfacetension\,\delta P/\Abub$.
Now, in the case of a pure shear $\Gamma$,
at least for a hexagonal, crystalline foam,
one can show that Eq.~(\ref{Eq:perimeter_dry})
yields $\delta P/P=\Gamma^2/8$.
Hence, $G=\frac14\,\surfacetension P/\Abub$
and $\dilatancy\simeq-\frac{\surfacetension\,P}{2\Abub}=-2G$.}.
For instance, in the case of a crystalline, 
hexagonal 2D foam: $\dilatancy\simeq-4\sqrt{3}\,\surfacetension/P$.

\mysection{Discussion}

In the present work,
we have shown that positive static dilatancy in liquid foams
might result from the overall increase in Plateau border lengths 
when the foam is being deformed elastically.
In 2D GG foams, this would be in fact the pseudo Plateau borders
(those along the solid walls).
The present study was restricted to crystalline foams,
whether 2D GG hexagonal foams or 3D Kelvin foams.
A weaker (or even reversed) effect
cannot be excluded, at this stage,
in the case of disordered or polydisperse foams,
or even simply for different 3D crystalline structures.
Indeed, in 3D at least, there seems to be no trustworthy reason
why the minimum in the total facette surface area
(which is the definition of a foam at rest, in the dry limit)
should coincide with the minimum in the sum of all Plateau border lengths.
For a GG foam in a somewhat less dry configuration
(regimes A-D of Fig.~\ref{Fig:diagram_pancake_regimes_ABCD}),
the expected dilatancy effect
can be shown to be less intense~\cite{dilatancy_geometry_article}.

The negative dilatancy predicted~\cite{weaire_2003_2747,rioual_2005_117}
in the dry limit in truly 2D foams
(invariant along the third dimension of space)
coincides with one particular limit of our study,
which indeed corresponds to a truly 2D foam.
It is then related to the interfacial (and deviatoric) part of the stress.
By contrast, in the other regimes, we obtain a positive dilatancy,
related to the pressure in both fluid phases.
Let us mention that
negative dilatancy is also possibly expected in the wet limit
of the pancake regime~\cite{dilatancy_geometry_article}.

If static dilatancy is confirmed experimentally,
it should be coupled with drainage models (which describe permeation) 
and incorporated into rheological models
of foams~\cite{labiausse_2007_479,marmottant_graner,marmottant_2008_EPJEII,saramito_2007_14,benito_2008_225_elasto_visco_plastic_foam}.

\mysection*{Acknowledgements}

We gratefully acknowledge fruitful discussions
with Benjamin Dollet, with Reinhard H\"ohler,
and with other participants of GDR 2983 Mousses (CNRS)
and of the Informal Workshop on Foam mechanics (Grenoble, 2008).
This work was supported by Agence Nationale de la Recherche 
(ANR-05-BLAN-0105-01).

\bibliography{letter_dilatancy}
\end{document}